# Tunable Magnonic Frequency and Damping in $[Co/Pd]_8$ Multilayers with Variable Co Layer Thickness

S. Pal[1], B. Rana[1], O. Hellwig[2], T. Thomson,[2] and A.Barman[1,a)]

[1]*Department of Material Sciences, S. N. Bose National Centre for Basic Sciences, Block JD, Sector III, Salt Lake, Kolkata 700 098, India*

[2]*San Jose Research Center, Hitachi Global Storage Technologies, 3403 Yerba Buena Rd., San Jose, California 95135, USA*

**Abstract**

We report the experimental observation of collective picosecond magnetization dynamics in $[Co/Pd]_8$ multilayers with perpendicular magnetic anisotropy. The precession frequency shows large and systematic variation from about 5 GHz to about 90 GHz with the decrease in the Co layer thickness from 1.0 nm to 0.22 nm due to the linear increase in the perpendicular magnetic anisotropy. The damping coefficient $\alpha$ is found to be inversely proportional to the Co layer thickness and a linear relation between the perpendicular magnetic anisotropy and $\alpha$ is established. We discuss the possible reasons behind the enhanced damping as the *d-d* hybridization at the interface and spin pumping. These observations are significant for the applications of these materials in spintronics and magnonic crystals.

[a)]Electronic mail: abarman@bose.res.in

Magnetic multilayers (ML) with perpendicular magnetic anisotropy (PMA) have attracted attention due to their potential applications in patterned magnetic media[1], spin transfer torque magnetic random access memory (STT-MRAM),[2-3] and magnonic crystals.[4] For applications in magnetic media and STT-MRAM devices, large precession frequency associated with the large PMA and a reliable and low damping constant $\alpha$ are desirable. On the other hand, for applications in magnonic crystals, broadly tunable magnonic frequencies and $\alpha$ with physical and material parameters are essential. All potential applications demand large and broadly tunable precession frequencies, small $\alpha$ values and a correlation between the PMA and $\alpha$. PMA is believed to originate from the interface anisotropy due to the broken symmetry and *d-d* hybridization[5] at the Co/Pd and Co/Pt interfaces. The competition between interface and volume anisotropies results in a variation of PMA with the thickness of the Co layer ($t_{Co}$) as has been reported in continuous[6] and patterned[7] magnetic multilayers. Consequently, a large variation in the precession frequency in the picosecond magnetization dynamics of these multilayers is expected. On the other hand, it has been predicted recently[8] that there may be a linear correlation between PMA and damping based on existing theoretical works.[9-10] The intrinsic Gilbert damping $\alpha$ and PMA both have their origins in the spin-orbit interaction and are approximately proportional to $\xi^2/W$, where $\xi$ is the spin-orbit interaction energy and $W$ is the *d*-band width. However, no clear correlation between the PMA and $\alpha$ has been observed so far.[8,11-12] Mizukami *et al.*[8] observed an increases in $\alpha$ with decrease in $t_{Co}$ but it was not inversely proportional to $t_{Co}$. In this work we studied the picosecond magnetization dynamics in $[Co(t_{Co})/Pd(0.9\ nm)]_8$ multilayers with $t_{Co}$ varying between 1.0 nm and 0.22 nm. We observed a systematic increase in the precession frequency and $\alpha$ with the decrease in $t_{Co}$. The extracted PMA, from the macrospin modeling of the precession frequency, shows a linear correlation with $\alpha$.

The ML structures are deposited by dc magnetron sputtering.[7] The base pressure of the deposition chamber was $2 \times 10^{-8}$ mbar and the deposition was performed at 3 mTorr of Ar pressure. A series of $[Co(t_{Co})/Pd(0.9\ nm)]_8$ stacks were prepared using a Ta(1.5 nm)/Pd(3.0 nm) seed layer, which ensured [111] textured ML with a mosaic spread of 7° full width at half maximum (FWHM). The Co layer thickness is varied from 0.13 nm to 1.0 nm in this experiment. The thickness of the Co and Pd layers were confirmed by x-ray reflectivity, and the average roughness at the interface was found to be about 0.05 nm. The magnetic hysteresis loops were measured by both polar magneto-optical Kerr effect (P-MOKE) and vibrating sample magnetometer (VSM) at room temperature. Figure 1(a) shows the magnetic anisotropy field ($H_K$) increases systematically with decrease in $t_{Co}$ and exhibits a maximum at 0.22 nm, beyond which it decreases sharply. The saturation magnetization ($M_S$), on the other hand, decreases monotonically with the decrease in $t_{Co}$ over the entire range. Figure 1(b) shows a typical square magnetic hysteresis loop for $t_{Co}$ = 0.36 nm. The ultrafast magnetization dynamics was probed by time-resolved magneto-optical Kerr effect (TR-MOKE) measurements in a two-color optical pump-probe setup.[11] The second harmonic ($\lambda$ = 400 nm) of a Ti-sapphire laser (pulse-width < 80 fs) was used to pump the samples, while the time-delayed fundamental ($\lambda$ = 800 nm) laser beam was used to probe the dynamics by measuring the Kerr rotation by a balanced photo-diode detector, which completely isolates the Kerr rotation and the reflectivity signals. The pump and the probe beams were focused and spatially overlapped onto the sample surface by a microscope objective with numerical aperture N. A. = 0.85 in a collinear geometry. A large magnetic field is first applied at a small angle (~ 10°) to the surface normal of the sample to saturate its magnetization. The magnetic field strength is then reduced to the bias field value ($H$), which ensures that the magnetization remains saturated along the bias field direction. The pump beam was chopped at 2 kHz frequency and a phase sensitive detection of the Kerr rotation was used. Figure 1(c) shows typical time-resolved reflectivity and the Kerr rotation data and the corresponding fast Fourier transform (FFT) spectra from the ML with $t_{Co} = 0.5$ nm at $H = 1.72$ kOe.

The precessional dynamics appears as an oscillatory signal above the slowly decaying part of the time-resolved Kerr rotation after a fast demagnetization within 500 fs, and a fast remagnetization within 10 ps. A bi-exponential background is subtracted before performing the FFT to find out the corresponding power spectra. The time-resolved reflectivity also shows an oscillation at about 77 GHz originating from thermally excited strain waves. The precessional frequency shows clear variation with the bias fields as opposed to the reflectivity signal. Figures 2(a)-(b) show the time-resolved Kerr rotations and the corresponding FFT spectra for six ML samples with $t_{Co} = 1.0$ nm, 0.75 nm, 0.5 nm, 0.36 nm, 0.28 nm and 0.22 nm at bias fields as shown in the figure. For samples with $t_{Co} < 0.22$ nm, no clear precession is observed. All samples show a single precession frequency due to the collective precession of the whole stack as if it is a single macrospin, which allows us to use the macrospin modeling of the Landau-Lifshitz-Gilbert equation[13] to analyze the frequency and damping. The variation of precession frequency with the bias field is plotted in Fig. 3(a) for various values of $t_{Co}$. The precession frequency increases sharply for $t_{Co} \leq 0.75$ nm indicating the sharp increase in the PMA in this range. For the experimental geometry, as shown in Fig. 3(b), the expression for precession frequency is

$$f = \frac{\gamma\mu_0}{\left(1+\alpha^2\right)}\left[H\frac{\sin\beta}{\sin\theta} + 2\frac{K_{eff}}{\mu_0 M_S} - M_S\right] \qquad [1],$$

where $\gamma$ is the gyromagnetic ratio, $\alpha$ is the damping coefficient, $\theta$ and $\beta$ are the angles made by the equilibrium magnetization ($M$) and the bias field ($H$) with x-axis and $K_{eff}$ is the effective magnetic anisotropy. $\theta$ is obtained by minimizing the total energy of the system, while $\beta$ is known from the experimental geometry. $\alpha$ is determined by fitting the time-resolved magnetization with a damped sine function

$$M(t) = M(0)e^{\frac{-t}{\tau}}\sin\left(\omega t - \phi\right) \qquad [2],$$

where $\tau = \frac{1}{2\pi f\alpha}$, $f$ is the experimentally obtained precession frequency and $\phi$ is the initial phase of oscillation.[14] The calculated frequencies are plotted as solid lines in Fig. 3(a) and are in good agreement with the experimental data. $\alpha$ is found to be inversely proportional to $t_{Co}$ over the entire range, as shown in Fig. 3(c). The extrapolation of the linear fit to $\alpha$ vs. $1/t_{Co}$ data upto $1/t_{Co} = 0$ gives $\alpha = 0.011$, which is comparable with the value for bulk Cobalt (0.01). In Fig. 3(d) we plot $K_{eff}$ and $M_S$ as a function of $t_{Co}$, as extracted from the macrospin modeling. $K_{eff}$ is also found to be inversely proportional to $t_{Co}$ similar to $\alpha$, indicating a clear linear correlation between $\alpha$ and $K_{eff}$. The values of $M_S$ obtained from the TR-MOKE measurements almost coincide with those obtained from the VSM loops. We have also calculated the variation of $M_S$ with $t_{Co}$ and found that consideration of slight magnetization of the Pd layers[15] (~ 210 emu/cc) gives good agreement between the experimental and the theoretical data.

In Fig. 4, we plot $\alpha$ as a function of $K_{eff}$, which clearly shows that $\alpha$ is directly proportional to $K_{eff}$ with a slope of $4.33\times10^{-8}$ cc/erg. The values of $\alpha$ reported here are lower than the previously published works.[8,11,16] Below, we discuss the possible mechanisms responsible for the enhancement of $\alpha$ from its intrinsic value. One common channel of dissipation of energy is by scattering of the uniform precession with short wavelength magnons due to the presence of inhomogeneities including defects, which should increase as the thickness reduces. However, it has been reported that for perpendicularly magnetized samples magnon scattering is less effective[17] and hence is ruled out in materials with high PMA. The second possibility is spin pumping,[18] caused by the spin current generated by the precession of magnetization of the Co layers entering into the Pd layers and getting absorbed due to its small spin diffusion length, thereby enhancing $\alpha$. This is usually accounted for by considering the variation of the relaxation frequency $G = \alpha\gamma M_S$ with $1/t_{Co}$.[8, 18] The slope of $G$ vs. $1/t_{Co}$ obtained in our case is only $3.2\times10^8$ rad/s as compared to the previously reported values of about $13\times10^8$ rad/s for $Pt/Ni_{80}Fe_{20}/Pt$ and $34\times10^8$ rad/s in Pt/Co/Pt films. The third possibility is the decrease in bandwidth $W$ of the Co atomic layer in contact with the Pd layer due to the Co 3*d*-Pd 5*d* hybridization,[5] This is primarily an interface effect and effectively increases both $\alpha$ and $K_{eff}$, as discussed earlier. The observation of direct proportionality between $\alpha$ and $K_{eff}$ strongly indicates that this may be the primary mechanism of enhancement of $\alpha$ in our experiment. The fourth possibility is the roughness and alloying effects at the interface.[8]

However, while interface roughness and alloying would increase $\alpha$, it would also decrease $K_{eff}$, which is opposite to our observation and hence this possibility is also ruled out. Other possibilities such as dephasing of multiple spin wave modes due to incoherent precession of the constituent layers and formation of perpendicular standing waves are negligible because of the observation of a collective precession of all the layers in the stack and uniform excitation of the whole stack, respectively.

In summary, we have studied the time-resolved magnetization dynamics in a series of $[Co(t_{Co})/Pd(0.9\ nm)]_8$ multilayers with variable Co layer thickness $t_{Co}$. The decrease in $t_{Co}$ increases the perpendicular magnetic anisotropy $K_{eff}$, which effectively increases the precession frequency and a broadly tunable precession frequency between about 5 GHz and 90 GHz is observed. The precession frequency was analyzed by macrospin modeling of LLG equation and the saturation magnetization $M_S$, $\alpha$, and $K_{eff}$ were independently obtained from the dynamics. Both $\alpha$ and $K_{eff}$ are inversely proportional to $t_{Co}$ and hence are directly proportional. The enhancement of $\alpha$ is possibly due to spin pumping and the *d-d* hybridization at the Co/Pd interfaces as both effects are inversely proportional to $t_{Co}$. However, only the later is directly correlated to the enhancement of $K_{eff}$ due to the decrease in bandwidth *W* of the Co atomic layer at the interface, while the former has no contribution to $K_{eff}$. Hence, we believe that in our case the enhancement of $\alpha$ is caused primarily due to the *d-d* hybridization effect. The observations of relatively low values of $\alpha$ associated with large $K_{eff}$ and their linear correlation are significant for their applications in the STT-MRAM devices and magnonic crystals.

We gratefully acknowledge the financial assistance from Department of Science and Technology, Govt. of India under the India-EU collaborative project "DYNAMAG" (grant number INT/EC/CMS (24/233552)) and the Nano Mission (grant number SR/NM/NS-09/2007).

**References:**

[1]T. Thomson, G. Hu, and B. D. Terris, Phys. Rev. Lett. **96**, 257204 (2006); O. Hellwig, A. Berger, T. Thomson, E. Dobisz, H. Yang, Z. Bandic, D. Kercher, and E. E. Fullerton, Appl. Phys. Lett. **90**, 162516 (2007).

[2]Y. Huai, F. Albert, P. Nguyen, M. Pakala, and T. Valet, Appl. Phys. Lett. **84**, 3118 (2004).

[3]S. Mangin, D. Ravelosona, J. A. Katine, M. J. Carey, B. D. Terris, and E. E. Fullerton, Nature Mater. **5**, 210 (2006).

[4]V. V. Kruglyak, S. O. Demokritov and D. Grundler, J. Phys. D **43**, 264001 (2010); S. Neusser and D. Grundler, Adv. Mater. **21**, 2927 (2009).

[5]N. Nakajima, T. Koide, T. Shidara, H. Miyauchi, H. Fukutani, A. Fujimori, K. Iio, T. Katayama, M. Nývlt, and Y. Suzuki, Phys. Rev. Lett. **81**, 5229 (1998).

[6] F. J. A. den Broeder, W. Hoving, and P. J. H. Bloemen, J. Magn. Magn. Mater. **93** 562 (1991).

[7]O. Hellwig, T. Hauet, T. Thomson, E. Dobisz, J. D. Risner-Jamtgaard, D. Yaney, B. D. Terris, and E. E. Fullerton, Appl. Phys. Lett. **95**, 232505 (2009).

[8]S. Mizukami, E. P. Sajitha, D. Watanabe, F. Wu, T. Miyazaki, H. Naganuma, M.Oogane, and Y. Ando, Appl. Phys. Lett. **96**, 152502 (2010).

[9]V. Kambersky, Czech J. Phys., Sect. B **26**, 1366 (1976).

[10]P. Bruno, *Physical Origins and Theoretical Models of Magnetic Anisotropy* (Ferienkurse des Forschungszentrums Jürich, Jürich, 1993).

[11]A. Barman, S. Wang, O. Hellwig, A. Berger, E. E. Fullerton, and H. Schmidt, J. Appl. Phys. **101**, 09D102 (2007).

[12]P. J. Metaxas, J. P. Jamet, A. Mougin, M. Cormier, J. Ferre, V. Baltz, B. Rodmacq, B. Dieny, and R. L. Stamps, Phys. Rev. Lett. **99**, 217208 (2007).

[13]L. D. Landau & E. Lifshitz, Phys. Z. Sowjetunion **8**, 153 (1935); T. L. Gilbert, Phys. Rev. **100**, 1243 (1955).

[14]A. Barman, S. Wang, J. Maas, A. R. Hawkins, S. Kwon, A. Liddle, J. Bokor, and H. Schmidt, Appl. Phys. Lett. **90**, 202504 (2007).

[15]P. F. Carcia, A. Suna, D. G. Onn, and R. van Antwerp, Superlatt. Microstruct. **1**, 101 (1985); F. J. A. den Broeder, H. C. Donkersloot, H. J. G. Draaisma, and W. J. M. De Jonge, J. Appl. Phys. **61**, 4317 (1937).

[16]R. Arias and D. L. Mills, Phys. Rev. B **60**, 7395 (1999); P. Landeros, R. E. Arias, and D. L. Mills, Phys. Rev. B **77**, 214405 (2008).

[17]G. Malinowski, K. C. Kuiper, R. Lavrijsen, H. J. M. Swagten, and B. Koopmans, Appl. Phys. Lett. **94**, 102501(2009).

[18]Y. Tserkovnyak, A. Brataas, and G. E. W. Bauer, Phys. Rev. Lett. **88**, 117601 (2002).

**Figure Captions:**

FIG. 1. (a) Dependence of magnetic anisotropy field ($H_K$) and the saturation magnetization ($M_S$) on the Co layer thickness $t_{Co}$ in $[Co/Pd]_8$ multilayer films, as measured by static magnetometry. (b) A typical hysteresis loop from the multilayer sample with $t_{Co}$ = 0.36 nm, obtained from polar MOKE measurement. (c) The time-resolved reflectivity and Kerr rotation signals and the corresponding FFT spectra from the multilayer sample with $t_{Co}$ = 0.5 nm, showing the frequencies of phonon and the precession of magnetization, respectively.

FIG. 2. (a) The time-resolved Kerr rotation data after a bi-exponential background subtraction and (b) the corresponding FFT spectra are shown for $[Co/Pd]_8$ films with different Co layer thickness $t_{Co}$. The solid lines in Fig. 2(a) correspond to the fit with Eq. [2]. The applied bias fields are also shown in the figure.

FIG. 3. (a) The bias field dependence of the experimental precession frequencies (symbols) and the calculated frequencies (solid line) with Eq. [1] are shown for multilayers with different Co layer thickness. (b) The geometry for the macrospin model is shown. (c) The damping coefficient $\alpha$ (symbols: experimental data, solid line: linear fit) is plotted as a function of $1/t_{Co}$. (d) The extracted perpendicular magnetic anisotropy $K_{eff}$ and the saturation magnetization $M_S$ (filled squares: from TR-MOKE, open circles: from static magnetometry) are plotted as a function of $t_{Co}$. The dashed line shows the calculated $M_S$ values, while the dotted line corresponds to the linear fit to $K_{eff}$ vs $t_{Co}$.

FIG. 4. The damping coefficient $\alpha$ is plotted as a function of $K_{eff}$ (symbols) and the dotted line corresponds to the linear fit.

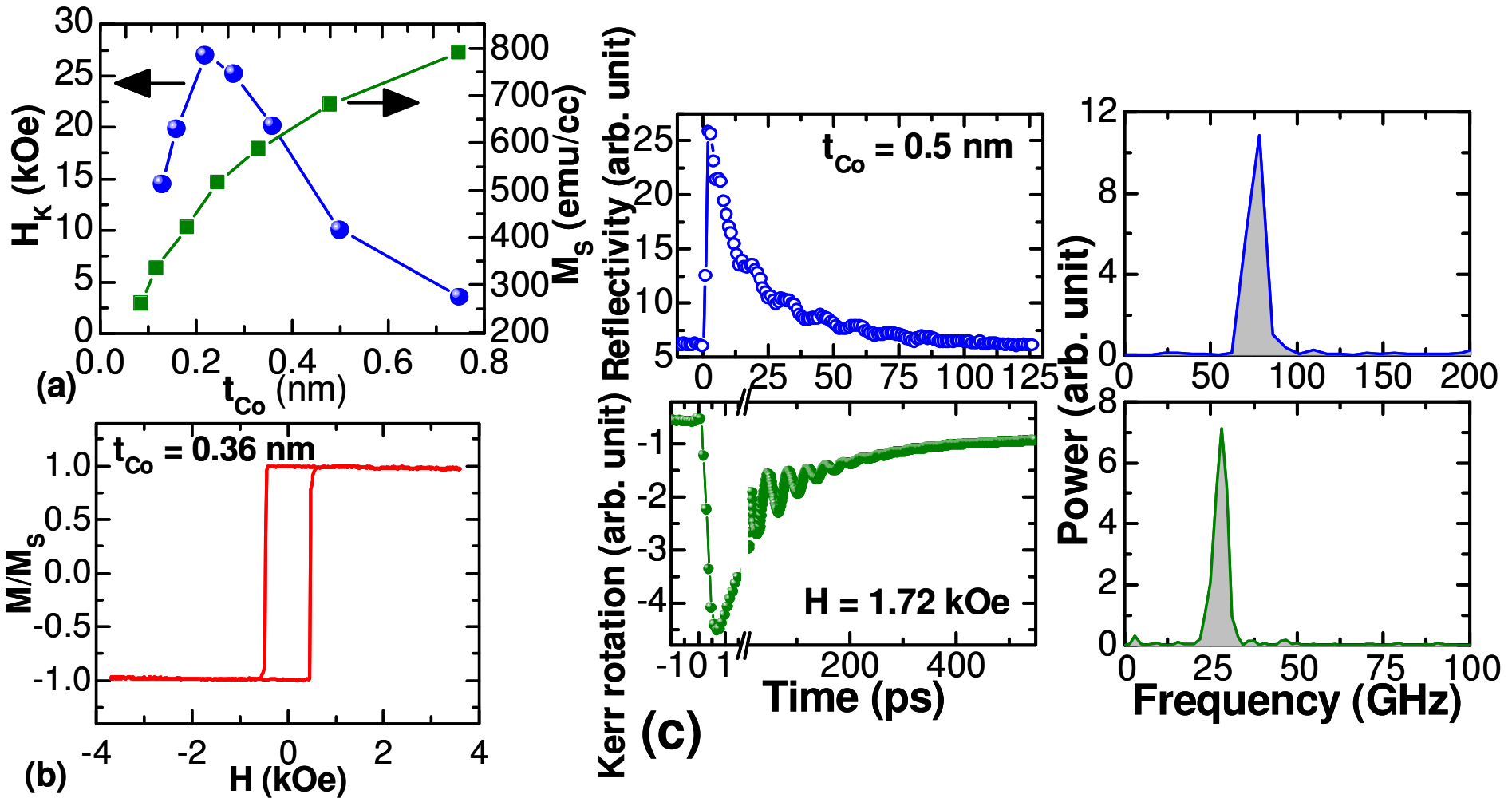

Fig. 1

S. Pal et al.

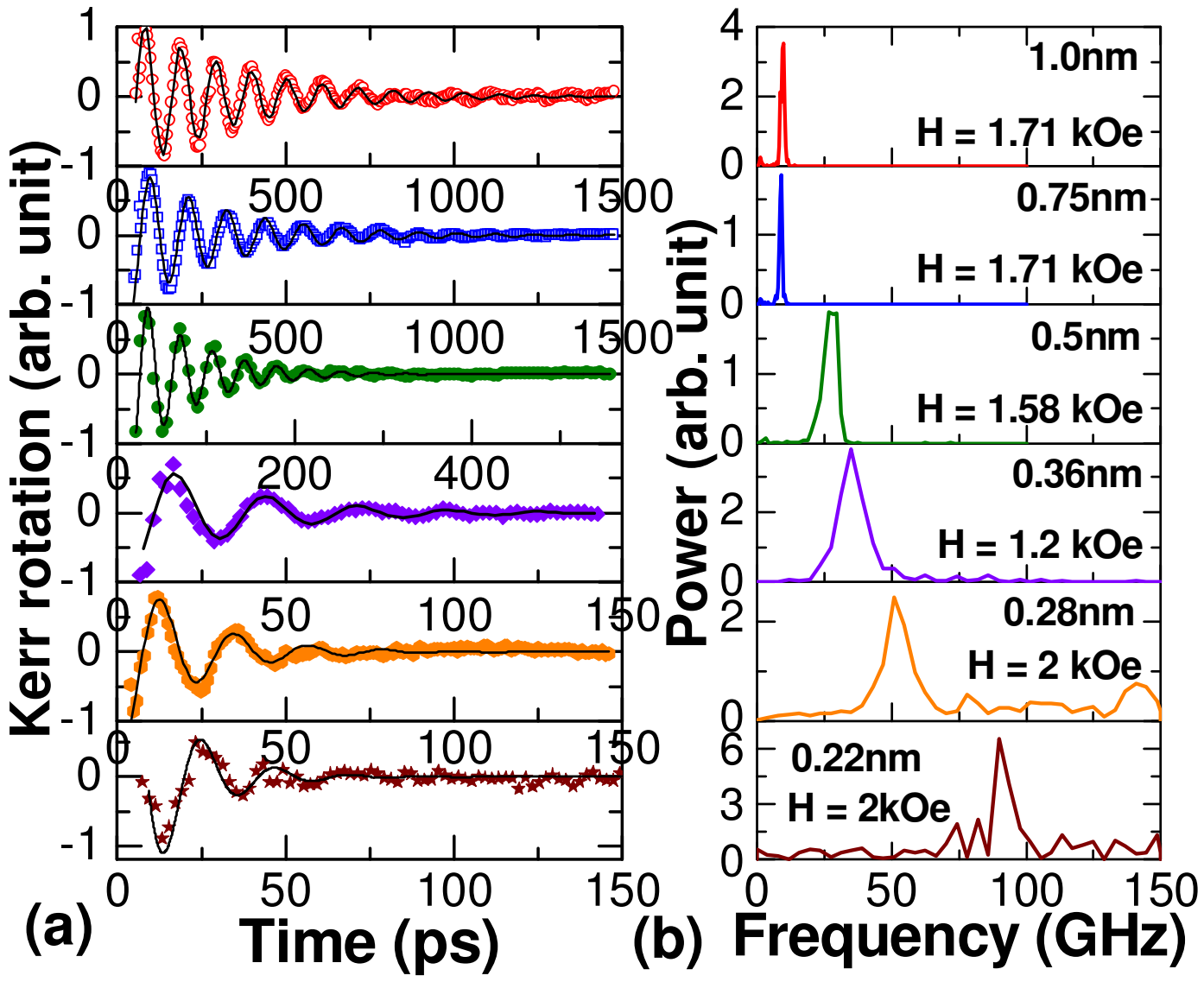

Fig. 2

S. Pal et al.

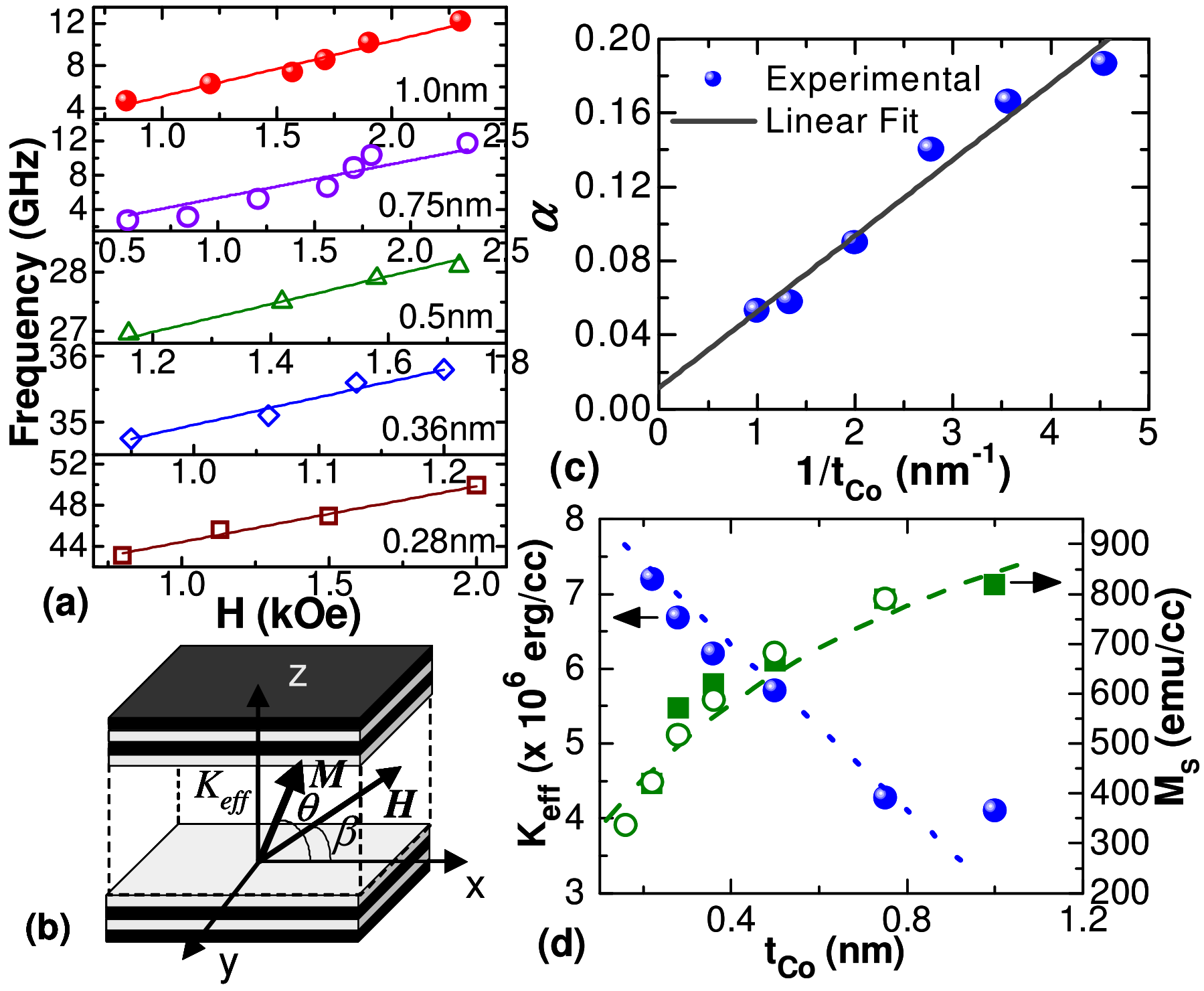

Fig. 3

S. Pal et al.

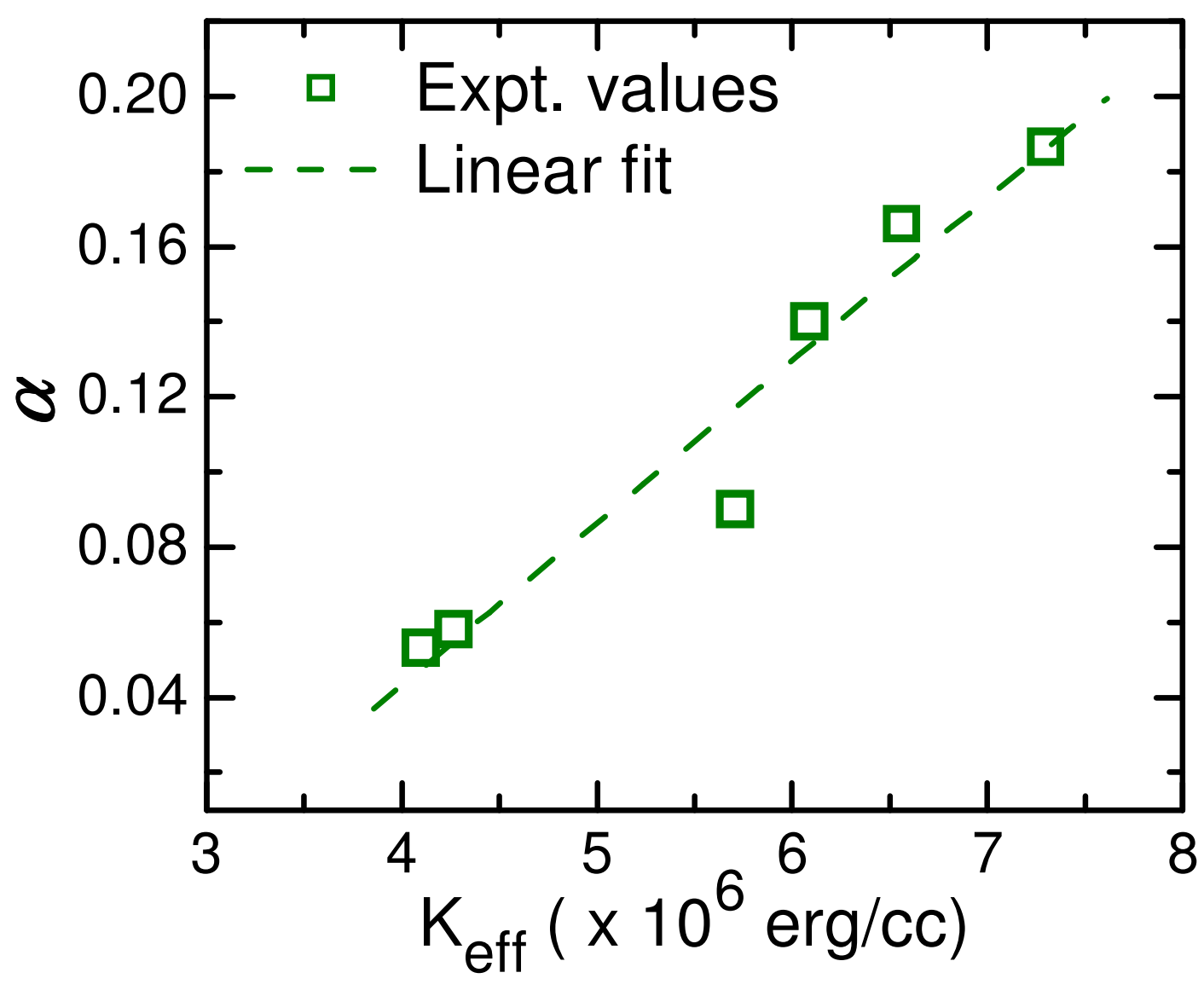

Fig. 4

S. Pal et al.